\documentclass[12pt]{article}
\usepackage[reqno]{amsmath}
\usepackage{epsfig}
\usepackage{array}
\usepackage{float}
\usepackage{lscape,graphicx}
\usepackage{graphics}
\usepackage{amssymb}
\usepackage{color}

\usepackage{array}
\newcolumntype{L}[1]{>{\raggedright\let\newline\\\arraybackslash\hspace{0pt}}p{#1}}
\newcolumntype{C}[1]{>{\centering\let\newline\\\arraybackslash\hspace{0pt}}p{#1}}
\newcolumntype{R}[1]{>{\raggedleft\let\newline\\\arraybackslash\hspace{0pt}}p{#1}}

\newcommand{\bea}{\begin{eqnarray}}
\newcommand{\eea}{\end{eqnarray}}
\newcommand{\ea}{\end{array}}

\newcommand{\be}{\begin{equation}}
\newcommand{\ee}{\end{equation}}
\newcommand{\bad}{\begin{array}{ccc}}

\newcommand{\ba}{\begin{array}{c}}
\begin{document}

\bibliographystyle{ieeetr}

\begin{center}
\mathversion{bold}
{\bf{\large Muon anomalous magnetic dipole moment in a low scale type I see-saw model}}
\mathversion{normal}

\vspace{0.4cm}
D. N. Dinh\\
%
\vspace{0.1cm}
{\em Institute of Physics, Vietnam Academy
of Science and Technology, \\
10 Dao Tan, Ba Dinh, Hanoi, Vietnam.\\
}
%
\end{center}
\begin{abstract}
Recent experimental results on muon anomalous magnetic dipole moment have shown a $4.2\sigma$ tension with the SM prediction, which has blown a fresh wind into the elementary particle physics community. The problem is believed to be explained only by physics beyond the standard model. Current work considers the anomalous moment in a scenario of models with mirror symmetry and type I see-saw mechanism at low energy scale of electroweak interactions. After a brief introduction to the model, a detailed numerical analysis of muon anomalous phenomenology will be carefully performed. Analysis results show that the model is not successful in explaining the muon moment problem, however the contributions of channels involving neutral Higgs scalars, including both the light and heavy ones, might provide sizable corrections to the discrepancy.
\end{abstract}

\section{Introduction}
	
In this letter, we are interested in the class of extended versions of the standard model with mirror symmetry and light active neutrino masses generated by the type I see-saw mechanism at the low energy scale of electroweak interactions \cite{Hung:2006ap}. A fermion mirror sector is proposed by introducing a corresponding mirror partner for each standard model fermion with the same quantum numbers but opposite chirality. The presence of mirror partners of left-handed neutrinos, thus the right-handed ones, provides a necessary condition for 
the type I see-saw mechanism to operate \cite{Minkowski:1977sc, Gell-Mann:1979vob, Yanagida:1979as, Mohapatra:1980yp}.  In contrast to the canonical type I see-saw, which operates at ultra-high energy scale, it is shown in that the new physics scale for the model under consideration might be as low as 100 GeV, and thus at the electroweak interaction scale.\\

We work on an updated version of the class of models that was introduced to accommodate the 125 GeV SM-like Higgs 
scalar discovery \cite{Hung:2015hra, Hoang:2014pda}. In contrast to the original ones, an additional Higgs doublet 
has been introduced, so there are two Higgs doublets, which are respectively responsible for mass generation in the 
normal and mirror sectors. Two candidates among the neutral scalars appearing after spontaneous symmetry breaking are shown to have
signals in agreement with ATLAS and CMS results \cite{Hoang:2014pda}. Besides, the model also has to confront the 
precision measurements of the electroweak processes, especially the effects of extra chiral doublets. A large parameter 
space is validated to be available after being constrained by EW precision data \cite{Hoang:2013jfa}. Moreover, the production of new fermions, 
whose masses are connected to the electroweak scale, and therefore can not exceed TeV, might be observable at the LHC with maximal 
running energy reaching 14 TeV. This appealing problem has been detailed discussed in \cite{Chakdar:2016adj}. The research showed 
that the $5\sigma$ discovery of $\nu_R$ is possible with as low as $20 fb^{-1}$ of integrated luminosity of the LHC running at 13 TeV   
for the RH neutrino mass within a range of $200-500$ GeV. For larger $\nu_R$ mass, up to $650$ GeV, the required luminosity might 
be as high as $100 fb^{-1}$. These conclusions are obtained based on an assumption of large Yukawa coupling $g_{\ell S}$ of the 
interaction between the normal lepton, mirror lepton and a light singlet scalar. However, this coupling should be much 
smaller due to being constrained by the current upper bounds on some lepton flavor violation processes 
\cite{Hung:2017voe, Dinh:2021fmt}, and also the muon anomalous magnetic moment as we will see latter in this research, therefore the real physics may happen in very different scenario and need a reconsideration \cite{Chakdar:2016adj}.\\

It is well-known recent time that the combination of Brookhaven E8211 result \cite{Muong-2:2006rrc} and $(g-2)_\mu$ experiment 
at Fermilab \cite{Muong-2:2021ojo}  for the muon anomalous magnetic moment has obtained the result
\begin{equation}
\Delta a_\mu=a_\mu^{EXP}-a_\mu^{SM}=(251\pm 59)\times 10^{-11},
\end{equation}
which is a $4.2\sigma$ discrepancy with the SM prediction \cite{Aoyama:2020ynm}. Although the result has not taken into account 
the recent lattice QCD calculations for hadronic vacuum polarization \cite{Borsanyi:2020mff} and the latest measurement of 
$e^+e^-\to {\rm hadrons}$ \cite{CMD-3:2023alj}, it can not be denied to be the strong evidence of new physics beyond the standard model.\\

From the theoretical perspective, a large number of researches has investigated the problem of muon anomalous 
magnetic dipole moment in various scenarios of physics beyond the standard model \cite{Popov:2016fzr, Lee:2022nqz, DelleRose:2020oaa, Babu:2022pdn, Kawamura:2022uft, Chowdhury:2022moc, Kim:2022hvh, Chowdhury:2022dps, Saez:2021qta, Chakrabarty:2022gqi, Ghorbani:2022muk, Cirigliano:2021peb}. 
In the class of models with mirror symmetry under consideration, the muon problem has also been briefly mentioned in some 
researches \cite{Hung:2015hra, Chang:2017vzi}, involving the channel with participation of light neutral scalar. However more detailed analysis should be performed, besides taking into account the contributions provided by other channels of heavy neutral and singly charged scalars.\\

This research discusses the phenomenology of muon anomalous magnetic dipole moment in the scenario of an extended version of the standard model with mirror symmetry, accommodating the 125 GeV SM-like scalar discovery. The contents are arranged as follows: besides the introduction in this section, in Sect. 2 we briefly introduce the model and required vertices for further calculations. In Sect. 3, we derive explicit form factors and algebraic expressions for muon anomalous magnetic dipole moment. Then, numerical analysis is also performed in this section. Finally, we give the conclusion in Sect. 4.   
\section{\label{model} A review of the model}
	
\subsection{The model content}

	This under consideration extended version of the EW-scale $\nu_R$ model is constructed based on the symmetric group $SU(2)\times U(1)_Y\times Z^4_{SM} \times Z^4_{MF}$, in which $SU(2)\times U(1)_Y$ is the gauge group, and $Z^4_{SM} \times Z^4_{MF}$  is a global symmetry introduced to forbid some unexpected interactions. The arrangement of scalar and matter fields under gauge group and their transformations under the global symmetry are detailed shown in the Table \ref{fieldTable}. Here, transformation of a given field $\Psi$ under the $Z^4_{SM} \times Z^4_{MF}$ is characterized by $(\omega_a^\alpha,\omega_b^\beta)$, which can be expressed as $\omega_a^\alpha\Psi\omega_b^\beta$, 
	where $\omega_a^4$=$\omega_b^4=1$.\\
	
 Note that the right-handed neutrinos in this model are components of $SU(2)\times U(1)_Y$ doublets; therefore they are non-sterile and take part in the weak interaction. Moreover, the heavy right-handed neutrinos naturally occur in the mirror sector accompanying the light active ones to fulfill the required conditions for the type I see-saw neutrino mass generation functioning. Five Higgs scalars (two doublets, two triplets and a singlet) are introduced to give masses for fermion particles, their roles could be seen in the later part of the paper.\\
    
\begin{table}[t]
\begin{center}
    \begin{tabular}{|c|c|c|c|}
    \hline
    Multiplets & $SU(2)\times U(1)_Y$ & $Z^4_{SM}$ & $Z^4_{MF}$ \\ \hline
    $\ell_L=(\nu_L,~e_L)^T$, $q_L=(u_L,d_L)^T$  & $(2,-1)$, $(2,1/3)$ & $\omega_a^3$ & 1 \\ 
    $\ell_R^M=(\nu_R,~e_R^M)^T$, $q_R^M=(u_R^M,~d_R^M)^T$ & $(2,-1)$, $(2,1/3)$ & 1 & $\omega_b^3$ \\ \hline
    $e_R$, $u_R$, $d_R$ & $(1,-2)$, $(1,4/3)$, $(1,-2/3)$ & $\omega_a$ & 1 \\ 
    $e_L^M$, $u_L^M$, $d_L^M$ & $(1,-2)$, $(1,4/3)$, $(1,-2/3)$ & 1 & $\omega_b$ \\ \hline
    $\Phi_2=(\phi_2^+,\phi_2^0)$ & (2,1) & $\omega_a^2$ & 1 \\ 
    $\Phi_{2M}=(\phi_{2M}^+,   \phi_{2M}^0)$ & (2,1) & 1 & $\omega_b^2$ \\ \hline
    $\chi=\left(\chi^{++},\chi^+,\chi^{0}\right)$ & (3,2) & 1 & $\omega_b^2$ \\ \hline
    $\xi=\left(\xi^+,\xi^0,\xi^{-}\right)$ & (3,0) & 1 & 1 \\ \hline
    $\phi_S$ & (1,0) & $\omega_a^3$ & $\omega_b$ \\ \hline
    \end{tabular}
    \end{center} 
    \caption{Model's field content and their transformations under gauge and global discrete symmetries, where $\omega_a^4$=$\omega_b^4=1$.} 
    \label{fieldTable} 
\end{table}
Before writing down the Yukawa couplings, let us briefly discuss the Higgs scalar sector. Apparently, the global symmetry defined in the earlier part only allows $\Phi_2$ to couple to SM fermions, while $\Phi_{2M}$ will couple to the mirror partners. The singlet 
$\phi_S$, which transforms nontrivially under both $Z^4_{SM}$ and $Z^4_{MF}$, will couple to a normal and a mirror fields. Finally, $\chi$ is responsible for introducing two unit lepton number violation term, which is needed for acquiring Majorana masses for heavy neutrinos. Detailed expressions of the Yukawa couplings to generate masses for matter fields are:
	\begin{equation}
	\label{YukawaLepton}
	 \mathcal{L}_Y^{\ell}=g_{\ell}\bar{\ell}_L\Phi_2 e_R+g_{\ell}^M\bar{\ell}_R^M\Phi_{2M} e_L^M
	 +g_{\ell s}\bar{\ell}_L\phi_s \ell_R^M+g_{\ell s}'\bar{e}_L^M\phi_s e_R + h.c.,
    \end{equation}	
    \begin{eqnarray}
	 \mathcal{L}_Y^{q}&=&g_u\bar{q}_L\tilde{\Phi}_2 u_R+g_{u}^M\bar{q}_R^M\tilde{\Phi}_{2M} u_L^M 
	 +g_d\bar{q}_L\Phi_2 d_R+g_{d}^M\bar{q}_R^M\Phi_{2M} d_L^M \nonumber\\
	 &&+g_{qs}\bar{q}_L\phi_s q_R^M+g_{u s}'\bar{u}_L^M\phi_s u_R+g_{d s}'\bar{d}_L^M\phi_s d_R + h.c.,
    \end{eqnarray}	
\begin{equation}
\label{LagMajoMass}
 \mathcal{L}_{\nu_R}=g_{M} \left(\ell_R^{M,T} \,\sigma_2\right)\, (i\tau_2\tilde{\chi}) \, \ell_R^{M} \, ,
\end{equation}    
where $\sigma_2$ is the second Pauli matrix, $\tilde{\Phi}_2=i\sigma_2\Phi_2^*$, $\tilde{\Phi}_{2M}=i\sigma_2\Phi_{2M}^*$, and $\tilde{\chi}$ form of the complex Higgs triplet with $Y=2$ is
	  \begin{equation}
	  \tilde{\chi}=\frac{1}{\sqrt{2}}\vec{\tau}.\vec{\chi}=
	  \left(\begin{array}{cc}
      \frac{1}{\sqrt{2}}\chi^+ & \chi^{++} \\
      \chi^0 & -\frac{1}{\sqrt{2}}\chi^+ 
       \end{array}\right).
	  \end{equation}  
Note that in the equation (\ref{LagMajoMass}), $\sigma_2$ acts on the space of two-component Weyl spinors, while $\tau_2$ acts on the SU(2) isospin space. We expect to have Hermetic charged fermion mass matrices, which simply implies $g_{\ell s}'=g_{\ell s}^\dagger$, 
$g_{u s}'=g_{u s}^\dagger$ and $g_{d s}'=g_{d s}^\dagger$, respectively. Finally, discussion on the quark sector will not be performed in this letter, because it is not involved in the phenomenology of physical 
quantity under consideration of this research.
\subsection{Symmetry breaking and mass generations}
We discuss in this subsection the mechanism of mass generations for fermion and scalar particles in this model, when the symmetry is spontaneously breaking. Let us suppose that Higgs fields develop their vacuum expectation values (VEV) as the following: 
$\langle\Phi_2\rangle=(0,v_2/\sqrt{2})^T$, $\langle\Phi_{2M}\rangle=(0,v_{2M}/\sqrt{2})^T$, 
$\langle\chi^0\rangle=v_M$, and $\langle\phi_S\rangle=v_S$.\\

The charged lepton mass matrix that can be easily obtained from eq. (\ref{YukawaLepton}), is explicitly expressed as
\begin{equation}
	\label{LeptonMass}
	M_\ell=\left( \begin{array}{cc}
	m_\ell & m_\ell^D \\
	(m_\ell^D)^\dagger & m_{\ell M}
	\end{array} \right) \,,
\end{equation}
where $m_\nu^D=m_\ell^D=g_{\ell s}v_S$, $m_\ell=g_{\ell} v_2/\sqrt{2}$, and 
$m_{\ell M}=g_{\ell}^M v_{2M}/\sqrt{2}$. Based on the current experimental status of searching new fermions beyond the standard model, one expects that masses of the mirror partners are much heavier than their normal ones, thus it is reasonable to assume 
$m_{\ell M}\gg m_\ell$ and $m_{\ell M},m_\ell\gg m_\ell^D$. This assumption allows us to approximately
block-diagonalize $M_\ell$ in the same way usually done for the see-saw type I neutrino mass matrix, then we obtain:
\begin{eqnarray}
\tilde{m}_\ell=m_\ell-\frac{(m_\ell^D)^2}{m_{\ell M}-m_{\ell}}\approx m_\ell,~~~
\tilde{m}_{\ell M}=m_{\ell M}+\frac{(m_{\ell M}^D)^2}{m_{\ell M}-m_{\ell}}\approx m_{\ell M},
\end{eqnarray}
\begin{equation}
	\left( \begin{array}{c}
	\ell_{L(R)} \\
	\ell^M_{L(R)}
	\end{array} \right) \,=\left( \begin{array}{cc}
	U_{\ell L(R)} & -R_\ell U_{\ell L(R)}^M \\
	R_\ell^\dagger U_{\ell L(R)} & U_{\ell L(R)}^M
	\end{array} \right) \,\left( \begin{array}{c}
	\ell'_{L(R)} \\
	\ell^{M'}_{L(R)}
	\end{array} \right),
	\label{LepBasisConver}
\end{equation}
where $\ell'_{L(R)}$, $\ell^{M'}_{L(R)}$ are respectively the normal and mirror charged leptons in the mass basis; $R_\ell\approx \frac{m_\ell^D}{ m_{\ell M}}\ll 1$, and  $\tilde{m}_\ell=U_{\ell L}
m_\ell^d U_{\ell R}^\dagger$, $\tilde{m}_{\ell M}=U_{\ell L}^Mm_{\ell M}^d {U_{\ell R}^M}^\dagger$,
in which $m_{\ell}^d$ and $m_{\ell M}^d$ are diagonal matrices.\\

After the gauge symmetry is spontaneously broken, the neutral leptons acquire their masses through a matrix of the canonical form of the type-I see-saw mechanism. By denoting that $M_R=g_M v_M$, one obtains
\begin{equation}
	\label{Mell}
	M_\nu=\left( \begin{array}{cc}
	0 & m_\nu^D \\
	(m_\nu^D)^T & M_{R}
	\end{array} \right) \,.
\end{equation}
Here, we have used the fact that the term of form $\left(\ell_L^{T} \,\sigma_2\right)\, (i\tau_2\tilde{\chi}) \, \ell_L$ is forbidden 
by the discrete symmetry $Z^4_{SM} \times Z^4_{MF}$ that forbids to generate Majorana mass term for the light active neutrinos at tree-level.
Moreover, it is shown in \cite{Hung:2006ap} that the mass term can arise at one-loop level, however its magnitude is about two order smaller
than that is generated by the typical see-saw mechanism, therefore the effects are reasonably ignored in this research.\\

Approximately block-diagonalizing (\ref{Mell}), while keeping in mind that
$M_{R}\gg m_\nu^D$, the result reads
\begin{equation}
\label{NeuMass}
\tilde{m}_\nu\approx -\frac{(m_\nu^D)^2}{M_{R}}=-\frac{(g_{\ell s} v_S)^2}{g_M v_M},~~ 
\tilde{m}_{\nu R}\approx M_{R}.
\end{equation}
We briefly comment on the light neutrino mass matrix $\tilde{m}_\nu$ defined in (\ref{NeuMass}), which is experimentally constrained to be at sub-eV order. 
In a canonical scenario, such a small constraint on $\tilde{m}_\nu$ implies $M_R$ should be very heavy $\sim 10^{9}$~GeV or higher. However, in the current model under consideration, if $(g^2_{\ell s}/g_M)\sim O(1)$ and $v_S\sim O(10^5~eV)$, $M_R$ can be much lower, at the electroweak scale. Note that this is the most interesting scenario that could be designed 
for this model to be testable at the LHC. However, $M_R$ is not forbidden to have larger values, up to about TeV, depending on both the magnitude of $v_M$ and interaction strength $g_M$. In fact, $g_{\ell s}$ is constrained by some
rare processes, for instance the $\mu\to e\gamma$ decay which has been detailed studied in \cite{Dinh:2021fmt}, that leads to $(g^2_{\ell s}/g_M)\ll 1$. In this case $v_S$ should be adjusted to be higher (might reach few GeV) to give correct masses for the light neutrinos if $M_R$ is fixed in range of hundred GeV.\\

Let $R_\nu\approx \frac{m_\nu^D}{ M_{ R}}$ be the ratio of the neutrino Dirac and Majorana mass matrices. Assuming that the light and heavy (mirror) neutrino mass matrices are diagonalized respectively by
$\tilde{m}_\nu=U_{\nu}^* m_\nu^d U_{\nu}^\dagger$, $\tilde{m}_{\nu R}={U_{\nu}^M}^*m_{\nu M}^d 
{U_{\nu}^M}^\dagger$, where $ m_\nu^d$ and $m_{\nu M}^d$ are diagonal matrices, we can obtain the  relations between the gauge and mass eigenstates of neutrinos as the following
\begin{equation}
	\left( \begin{array}{c}
	\nu_{L} \\
	(\nu_{R})^c
	\end{array} \right) \,=\left( \begin{array}{cc}
	U_{\nu} & -R_\nu U_{\nu}^M \\
	R_\nu^\dagger U_{\nu} & U_{\nu}^M
	\end{array} \right) \,\left( \begin{array}{c}
	\chi_L \\
	\chi_L^M
	\end{array} \right).
	\label{NuBasisConver}
\end{equation}

Before we introduce the masses and mass states of the new physical scalars that appear after spontaneous symmetry breaking, let us briefly explain why we need two Higgs triplets in this model.
It is well known that when triplets are introduced, the tree-level result $\rho=1$, which is precisely measured by experiment, will be violated. Fortunately, it is also shown in \cite{Chanowitz:1985ug} that, in a scenario of two triplets with appropriate hyper-charges, they can combine to form a $(3,3)$ representation under the global $SU(2)_L\otimes SU(2)_R$ symmetry. After symmetry breaking, the custodial $SU(2)$ symmetry is preserved and $\rho=1$. Thus, along with the triplet $\tilde{\chi}$, we need to add a real Higgs triplet with $Y=0$, denoted by $\left(\xi^+,\xi^0,\xi^{-}\right)$. 
The combinations of two triplets $(3,3)$ and two doublets $(2,2)$ under the global symmetry can be respectively expressed as:
\begin{equation}
	\label{chi}
	\chi = \left( \begin{array}{ccc}
	\chi^{0} &\xi^{+}& \chi^{++} \\
	\chi^{-} &\xi^{0}&\chi^{+} \\
	\chi^{--}&\xi^{-}& \chi^{0*}
	\end{array} \right) \,,
\end{equation}
\begin{equation}
\Phi_2= \left( \begin{array}{cc}
	\phi_2^{0,*} & \phi_2^+ \\
	\phi_2^- & \phi_2^{0}
	\end{array} \right),~~~
	\Phi_{2M}= \left( \begin{array}{cc}
	\phi_{2M}^{0,*} & \phi_{2M}^+ \\
	\phi_{2M}^- & \phi_{2M}^{0}
	\end{array} \right).
\end{equation}
Then the proper vacuum alignment for breaking gauge symmetry  from $SU(2)_L \times U(1)_Y$ to $U(1)_{em}$ can be easily written down:
\begin{equation}
\label{chivev}
\langle \chi \rangle = \left( \begin{array}{ccc}
v_M &0&0 \\
0&v_M&0 \\
0&0&v_M
\end{array} \right) \,,
\end{equation}
\begin{equation}
\label{DoubletVEV}
\langle\Phi_2\rangle= \left( \begin{array}{cc}
	v_2/\sqrt{2} & 0 \\
	0 & v_2/\sqrt{2}
	\end{array} \right),~~~
	\langle\Phi_{2M}\rangle= \left( \begin{array}{cc}
	v_{2M}/\sqrt{2} & 0 \\
	0 & v_{2M}/\sqrt{2}
	\end{array} \right).
\end{equation}
Due to the experimental constraint on the $W_\mu$ mass, VEVs of the real components of $\Phi_2$, $\Phi_{2M}$ and $\chi$ denoted in (\ref{chivev}) and (\ref{DoubletVEV}),  satisfy the conditions:
\be
	v_2^2 + v_{2M}^2 + 8\,v_M^2 = v^2\,,
\ee
where $v \approx 246 ~GeV$. Related to the above VEVs and for further discussions, the following notations are used: 
\be
\label{eq:sdefs}
	s_2 = \dfrac{v_2}{v}; ~~s_{2M} = \dfrac{v_{2M}}{v}; ~~s_M = \dfrac{2 \sqrt{2}\; v_M}{v}\,.
\ee
We analyze the physical scalar spectrum in this model after the gauge symmetry and the L-R global symmetry of the Higgs potential are spontaneously broken to the custodial $SU(2)_D$. Out of the seventeen degrees of freedom of the two Higgs triplets (one real and one complex) and two Higgs doublets, three of them are eaten to give masses to W’s and Z, while the rest are rearranged to form new physical Higgs bosons. Those that are mass-degenerate are grouped in the same physical scalar multiplets of the global custodial symmetry. Thus we have a five-plet (quintet) $(H_5^{\pm\pm},\;H_5^\pm,\;H_5^0)$, two triplets $(H_{3}^\pm,\;H_{3}^0)$, $(H_{3M}^\pm,\;H_{3M}^0)$ and three singlets $H_1^0,\; H_{1M}^0,\; H_1^{0\prime}\,$.\\

We will not introduce any specific case of Higgs potential in this research, but the above discussion on physical scalars applies to any one of them that possesses $SU(2)_L\otimes SU(2)_R$ global symmetry, including the cases that have been considered in detail in \cite{Hoang:2014pda}. It is reasonable to assume that the scalars mentioned earlier have masses at the electroweak scale, in the range of hundred to few hundred GeVs, because they are remnants of the gauge symmetry breaking mechanism. Recall that the scalars that are members of a multiplet (not singlet) have the same masses, while the three singlets $H_1^0,\; H_{1M}^0,\; H_1^{0\prime}\,$ are not physical states, in general. These gauge states are linear combinations of some mass eigenstates $(\tilde{H}_1^0,\; \tilde{H}_2^0,\; \tilde{H}_3^0\,$) as $H_1^0=\sum_{i}^3\alpha_i\tilde{H}_i$, $H_{1M}^0=\sum_{i}^3\alpha_i^M\tilde{H}_i$, where $\sum_{i}^3|\alpha_i|^2=1$ and $\sum_{i}^3|\alpha_i^M|^2=1$. The SM-like Higgs scalar discovered by LHC with mass 125-GeV is one of these three mass states \cite{Hoang:2014pda}.\\

Finally, let us discuss $\phi_s^0$, which is the remaining degree of freedom of the Higgs sector, originating from $\phi_S$. As a singlet, $\phi_S$ does not participate in the gauge symmetry breaking mechanism, so its VEV $v_S$ can have a large range of values from keV to few GeV. In this research, we will consider $\phi_s^0$ mass to have the same order of magnitude as $v_S$.
	
\subsection{The LFV vertices}
In the SM, there is no flavor changing of the neutral current at tree-level in the lepton sector, so there is no LFV vertex, because the charged lepton mass matrix and the matrix of Yukawa couplings are simultaneously diagonal, and the vector gauge bosons only interact with the left-handed components of the matter fields. However, these properties do not hold in this model, because the matter content has been enlarged with mirror fermions, and the vector fields also interact with the right-handed components of the mirror sector. Therefore, LFV interactions occur at tree-level for both the charged currents and Yukawa couplings. Their detailed expressions in the gauge basis can be found in \cite{Hoang:2014pda}.\\

To facilitate further discussion, we present below the LFV couplings in this model in the mass eigenstate basis. For consistency with the current experimental observations and for simplicity, we assume that the charged lepton and mirror charged lepton mixing matrices are real (so all the involved complex phases are neglected) and $U_{\ell L}=U_{\ell R}=U_{\ell}$, $U_{\ell L}^M=U_{\ell R}^M=U_{\ell}^M$. After dropping terms that are subleading of the second order of $R_{\nu(\ell)}$ and higher, the LFV couplings are given in table \ref{TableVertex} and eqs. from (\ref{YuV01}) to (\ref{YuV06}):
\begin{table}[h]
\centering
	\scalebox{1}{
		\begin{tabular}{|c|c|}
			\hline
			Vertices & Couplings\\
			\hline
			$(\bar{e}'_L\gamma^\mu \chi_L)W_\mu^-$ & $ -i\frac{g}{\sqrt{2} }U_{W_\mu}^{L}= -i\frac{g}{\sqrt{2} }U_{PMNS}$
			\\
			\hline 
			$(\bar{e}'_L\gamma^\mu \chi_L^M)W_\mu^-$ & $  -i\frac{g}{\sqrt{2}}U_{W_\mu}^{ML}= i\frac{g}{\sqrt{2} }
			\tilde{R}_\nu \left(U_{PMNS}^{M}\right)^*$ 
			 \\
			\hline
			$(\bar{e}'_R\gamma^\mu \chi_L^c)W_\mu^-$ & $  -i\frac{g}{\sqrt{2}}U_{W_\mu}^{R}= -i\frac{g}{\sqrt{2} }
			\tilde{R}_\nu^T \left(U_{PMNS}\right)^*$
			 \\
			\hline
			$\bar{e}'_R\chi_L  H_3^-$ & $  -i\frac{g}{2}Y_{H_3^-}^{L}= -i\frac{g~s_M}{2M_W c_M}m_{\ell}^dU_{PMNS}$ 
			\\
			\hline 
			$\bar{e}'_R\chi_L^M  H_3^-$ & $  -i\frac{g}{2}Y_{H_3^-}^{ML}= i\frac{g~s_M}{2M_W c_{M}}m_{\ell}^d
            \tilde{R}_\nu \left(U_{PMNS}^{M}\right)^*$ 
			 \\
			\hline
			$\bar{e}'_L\chi_L^{Mc}  H_3^-$ & $  -i\frac{g}{2}Y_{H_3^-}^{MR}= -i\frac{g~s_M}{2M_W c_M}
			\tilde{R}_\ell m_{\ell M}^dU_{PMNS}^M$ 
			 \\
			\hline
			$\bar{e}'_R\chi_L  H_{3M}^-$ & $  -i\frac{g}{2}Y_{H_{3M}^-}^{L}= -i\frac{g~s_{2M}}
			{2M_W s_2 c_M}m_{\ell}^dU_{PMNS}$ 
			\\
			\hline 
			$\bar{e}'_R\chi_L^M  H_{3M}^-$ & $  -i\frac{g}{2}Y_{H_{3M}^-}^{ML}= i\frac{g~s_{2M}}
			{2M_W s_2 c_M}m_{\ell}^d\tilde{R}_\nu \left(U_{PMNS}^{M}\right)^*$ 
			 \\
			\hline
			$\bar{e}'_L\chi_L^{Mc}  H_{3M}^-$ & $  -i\frac{g}{2}Y_{H_{3M}^-}^{MR}= -i\frac{g~s_M}{2M_W s_{2M} c_{M}}
			\tilde{R}_\ell m_{\ell M}^dU_{PMNS}^M$ 
			\\
		   \hline
			$\bar{e}'_R {e_L^M}' \phi^0_s$ & $  -i\frac{g}{2}Y_{\phi^0_s}^{ML}= -iU^{\dagger}_{\ell R}
			g_{\ell s}U_{\ell L}^M=-i \tilde{g}_{\ell s}$ 
			\\
			\hline 
			$\bar{e}'_L {e_R^M}' \phi^0_s$ & $  -i\frac{g}{2}Y_{\phi^0_s}^{MR}= -i~U^{\dagger}_{\ell L}
			g_{\ell s}U_{\ell R}^M=-i \tilde{g}_{\ell s}$ 
			\\
			\hline 
	\end{tabular}}
	\caption{Vertices involving muon anomalous magnetic dipole moment in the mass eigenstate basis. }
	\label{TableVertex}
\end{table}

\begin{equation}
\label{YuV01}
(\bar{e}'_R {e_L^M}' {\tilde{H}}_i^0)~~~~~~-i\frac{g}{2}Y^{ML}_{{\tilde{H}}_i^0}=
-i\frac{g}{2M_W}\left[\frac{\alpha_i}{s_2}m_\ell^d\tilde{R}_\ell
+\frac{\alpha_i^M}{s_{2M}}\tilde{R}_\ell m_{\ell M}^d\right],
\end{equation}
\begin{equation}
(\bar{e}'_L {e_R^M}' {\tilde{H}}_i^0)~~~~~~
-i\frac{g}{2}Y^{MR}_{{\tilde{H}}_i^0}=-i\frac{g}{2M_W}\left[\frac{\alpha_i}{s_2}m_\ell^d\tilde{R}_\ell
+\frac{\alpha_i^M}{s_{2M}}\tilde{R}_\ell m_{\ell M}^d\right],
\end{equation}

\begin{equation}
(\bar{e}'_R {e_L^M}' {H}_3^0)~~~~~~-i\frac{g}{2}Y^{ML}_{{H}_3^0}=-i\frac{g}{2M_W}\left[\frac{s_M}{c_M}m_\ell^d\tilde{R}_\ell
+\frac{s_M}{c_M}\tilde{R}_\ell m_{\ell M}^d\right],
\end{equation}
\begin{equation}
(\bar{e}'_L {e_R^M}' {H}_3^0)~~~~~~-i\frac{g}{2}Y^{MR}_{{H}_3^0}=-i\frac{g}{2M_W}\left[-\frac{s_M}{c_M}m_\ell^d\tilde{R}_\ell
-\frac{s_M}{c_M}\tilde{R}_\ell m_{\ell M}^d\right],
\end{equation}

\begin{equation}
(\bar{e}'_R {e_L^M}' {H}_{3M}^0)~~~~~~-i\frac{g}{2}Y^{ML}_{{H}_{3M}^0}=-i\frac{g}{2M_W}\left[-\frac{s_{2M}}{s_2}m_\ell^d\tilde{R}_\ell
-\frac{s_2}{s_{2M}}\tilde{R}_\ell m_{\ell M}^d\right],
\end{equation}
\begin{equation}
\label{YuV06}
(\bar{e}'_L {e_R^M}' {H}_{3M}^0)~~~~~~-i\frac{g}{2}Y^{MR}_{{H}_{3M}^0}=-i\frac{g}{2M_W}\left[\frac{s_{2M}}{s_2}m_\ell^d\tilde{R}_\ell
+\frac{s_2}{s_{2M}}\tilde{R}_\ell m_{\ell M}^d\right].
\end{equation}
Here one has used the notations $U_{PMNS}=U_\ell^\dagger U_\nu$, which is the famous PMNS mixing matrix, $U_{PMNS}^M=U_\ell^{M\dagger} U_\nu^M$ and $\tilde{R}_{\ell (\nu)}=U_{\ell}^\dagger R_{\ell (\nu)}U_{\ell}^M$. 

\mathversion{bold}
\section{Phenomenology of muon anomalous magnetic dipole moment} 
\subsection{One-loop form-factors and muon anomalous magnetic dipole moment}
\mathversion{normal}
In this scenario, under the considered model, the muon anomalous magnetic moment and the $\mu\to e +\gamma$ decay rate are related to the loop integral factors. In some previous research, one-loop diagrams with various kinds of internal lines have been calculated \cite{Dinh:2021fmt, Alonso:2012ji, Akeroyd:2009nu, Dinh:2012bp, Dinh:2019jdg}. The current work takes into account the effective charged lepton flavor-changing operators arising at one-loop, where the virtual particles running inside are either physical Higgs scalars (which include single and neutral charges, heavy and light ones) or W gauge bosons accompanied by relevant leptons. One might wonder on the contribution of doubly charged Higgs scalar channel. However it can be easily estimated to give sub-leading contribution due to the fact that the complex Higgs triplet with $Y=2$ can only interact with the lepton mirror doublet; therefore the channels involving doubly 
charged Higgs scalar can provide only high order corrections to the anomalous magnetic dipole moments of the normal charged leptons. Feynman diagrams of the dominated channels are shown in Fig.\ref{FeynDiagram}. The calculation result can be summarized as follows:
\begin{figure}[t]
\begin{center}
\includegraphics[width=15cm,height=5.5cm]{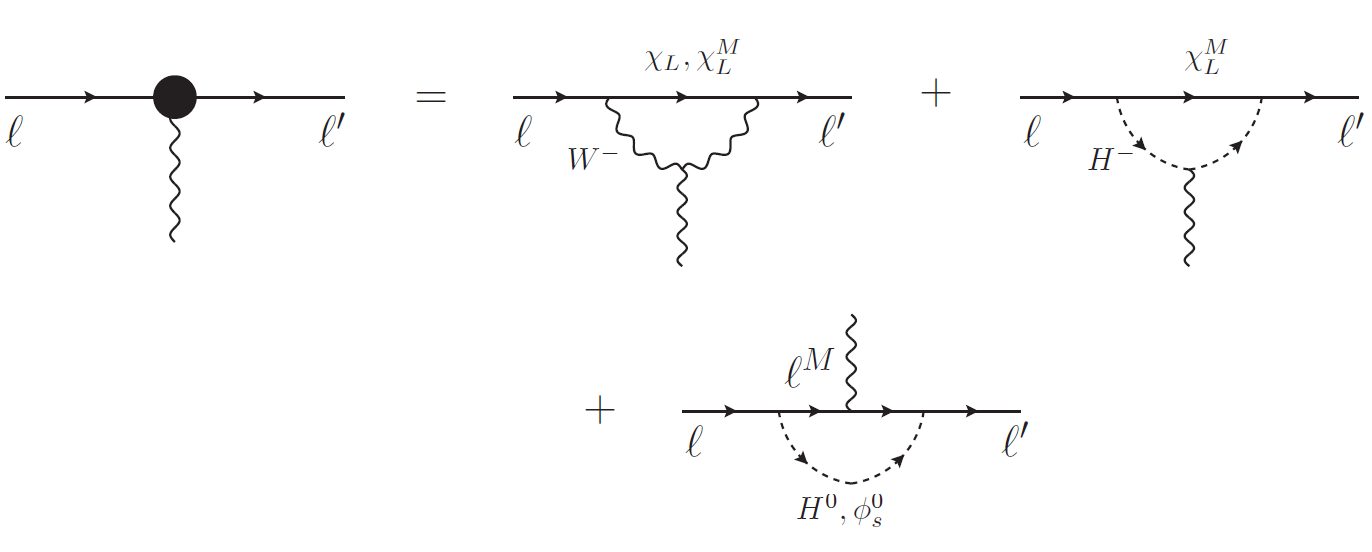}
\caption{Three group of diagrams give leading contributions to the muon anomalous magnetic dipole moment. Here, $H^-$ stands for $H_3^-$ and $H_{3M}^-$;
$H^0$ stands for ${\tilde{H}}_i^0$, $H^0_3$ and $H^0_{3M}$. 
}
\label{FeynDiagram}
\end{center}
\end{figure}
%
\begin{equation}
\label{effectiveLag}
\mathcal{L}_{eff}=-4\frac{eG_F}{\sqrt{2}}\left[ (m_\ell A_R+m_{\ell'}A_L)\bar{\ell'}\sigma_{\mu\nu}P_R\ell
+(m_\ell A_L+m_{\ell'}A_R)\bar{\ell'}\sigma_{\mu\nu}P_L\ell\right] F^{\mu\nu}.
\end{equation}
Here $A_{L,R}$ are the form factors:
\begin{eqnarray}
\nonumber
A_R=&&-\sum_{{H^Q},k}\frac{M_W^2}{64\pi^2M_H^2}\left[\left(Y_H^L\right)_{\mu k}\left(Y_H^L\right)_{e k}^*G_H^Q(\lambda_k)
+\frac{m_k}{m_\mu}\left(Y_H^R\right)_{\mu k}\left(Y_H^L\right)_{e k}^*\times R_H^Q(\lambda_k)\right]\\
&&+\frac{1}{32\pi^2}\sum_{k}\left[\left(U_{W_\mu}^L\right)_{\mu k}\left(U_{W_\mu}^L\right)_{e k}^*
G_{\gamma}(\lambda_k)-\left(U_{W_\mu}^R\right)_{\mu k}\left(U_{W_\mu}^L\right)_{e k}^*
\frac{m_k}{m_\mu}R_{\gamma}(\lambda_k)\right],\label{FunAR}
\end{eqnarray}
\begin{eqnarray}
\nonumber
A_L=&&-\sum_{{H^Q},k}\frac{M_W^2}{64\pi^2M_H^2}\left[\left(Y_H^R\right)_{\mu k}\left(Y_H^R\right)_{e k}^*G_H^Q(\lambda_k)
+\frac{m_k}{m_\mu}\left(Y_H^L\right)_{\mu k}\left(Y_H^R\right)_{e k}^*R_H^Q(\lambda_k)\right]\\
&&+\frac{1}{32\pi^2}\sum_{k}\left[\left(U_{W_\mu}^R\right)_{\mu k}\left(U_{W_\mu}^R\right)_{e k}^*
G_{\gamma}(\lambda_k)-\left(U_{W_\mu}^L\right)_{\mu k}\left(U_{W_\mu}^R\right)_{e k}^*
\frac{m_k}{m_\mu}R_{\gamma}(\lambda_k)\right],\label{FunAL}
\end{eqnarray} 
where $H^Q=\phi_S^0, \tilde{H}^{0}_{i}$ $(i=1,2,3)$, $H^{0}_{3}, H^{0}_{3M}$, $H^{+}_{3}, H^{+}_{3M}$, and
$m_k$ are the masses of associated fermions that accompany with either $H^Q$ or $W_\mu$ in the loops. The functions 
$G_{H}^Q(x)$, $R_{H}^Q(x)$, $G_{\gamma}(x)$, and $R_{\gamma}(x)$ appearing in eqs. (\ref{FunAR}) and (\ref{FunAL}) 
are defined as:
\begin{eqnarray}
G_H^Q(x)&=&-\frac{(3Q-1)x^2+5x-3Q+2}{12(x-1)^3}+\frac{1}{2}\frac{x(Qx-Q+1)}{2(x-1)^4}\log(x),\\
R_H^Q(x) &=& \frac{(2Q-1)x^2-4(Q-1)x+2Q-3}{2(x-1)^3}-\frac{Qx-(Q-1)}{(x-1)^3}\log(x),~~~~~~\\
G_\gamma(x)&=&\frac{10-43x+78x^2-49x^3+4x^4+18x^3\log(x)}{12(x-1)^4},
\\
R_\gamma(x)&=&-\frac{x^2+x-8}{2(x-1)^2}+\frac{3x(x-2)}{(x-1)^3}\log(x),
\end{eqnarray}
where $\lambda_k=m_k^2/M_{{W_\mu}({H^Q})}^2$ has been denoted.\\

Note that functions introduced in the above equations $G_H^Q(x)$, $R_H^Q(x)$, $G_\gamma(x)$, and $R_\gamma(x)$ are valid for
$x$ variable varying in interval $[0,+\infty)$, and get finite values at the specific points, such as $x=0,~1$ or when $x$ tends to infinity. Compare with the original expression, introduced in an earlier publication \cite{Dinh:2012bp}, $G_\gamma(x)$ has been divided by $4$ to be consistent to the factor $1/(32\pi^2)$ in the definitions of $A_L$ and $A_R$. The two functions, $G_H^Q(x)$ and $R_H^Q(x)$, to the best of the author's knowledge, have not been used in any other discussion on the
muon anomalous dipole moment so far. \\

The expression for muon anomalous magnetic dipole moment can be easily extracted from the effective Lagrangian (\ref{effectiveLag}), 
which arrives at   
\begin{equation}
\label{muonMDM}
\Delta a_\mu=\frac{4\pi\alpha_{em}}{\sin^2\theta_w}\frac{m_\mu^2}{M_W^2}(A_L+A_R),
\end{equation}
where $\alpha_{em}=1/137$ is the fine-structure constant. Note that formula (\ref{muonMDM}) should not include the contributions of light neutrino and $W_\mu$ loops, which have been taken into account in the standard model.
\mathversion{bold}
\subsection{Numerical analysis of muon anomalous magnetic dipole moment}
\mathversion{normal}
In this section, we perform numerical analysis of the muon anomalous magnetic dipole moment expressed by (\ref{muonMDM}) using current experimental data. To better understand the role of each kind of diagram and for convenience, we separately consider the contributions of one-loop diagrams with virtual $W$ gauge boson, neutral and singly charged Higgs scalars to the quantity. Moreover, for simplicity in further numerical discussions, we assume that three heavy neutrinos are degenerated in masses, which are denoted as $m_\chi^M$. Similarly, we make the same assumption for three mirror charged lepton masses $m_\ell^M$.\\

Before performing a detail discussion, let's make a rough estimation of the 
magnitudes of $R_{\nu(\ell)}$, which have central roles in the phenomenology of lepton flavour violation and muon anomalous magnetic dipole moment in this model. Starting from the expression of light neutrino mass matrix in (\ref{NeuMass}) and its current experimental constraint, we can easily obtain
\begin{eqnarray} 
R_\nu=\frac{m_\nu^D}{M_R}\sim 10^{-5}\sqrt{\frac{1{\rm GeV}}{M_R}}. 
\end{eqnarray}
For $M_R\sim 100$ GeV, $|R_\nu|$ has value at order of $10^{-6}$. The same
magnitude $|R_\ell|\sim 10^{-6}$ is analogously obtained if mirror charged lepton mass matrix is supposed not to be larger than the EW scale. Note that it is also relevant to estimate $|\tilde{R}_{\ell}|=|U_{\ell}^\dagger R_{\ell}U_{\ell}^M|$, as well as $|\tilde{R}_{\nu}|=|U_{\ell}^\dagger R_{\nu}U_{\ell}^M|$, to be at the same order as $|R_{\ell(\nu)}|\sim 10^{-6}$, due to the basis transformation matrices $U_{\ell}$ and $U_{\ell}^M$ are normalized.\\

Before carrying on discussions that are distinctive for the current model, let’s re-obtain the contribution of light active neutrinos and W-boson 1-loop diagrams by applying eq. (\ref{muonMDM}) for the corresponding interacting couplings while keeping in mind the unitarity of PMNS matrix. The outcome arrives at
\begin{equation}
\label{LigtNeutrino}
\Delta a_\mu^{\rm SM}({\chi_L})=\frac{\alpha}{8\pi\sin^2\theta_w}\frac{m_\mu^2}{M_W^2}G_\gamma(\lambda_{\chi_L})=\frac{G_F m_\mu^2}{4\sqrt{2}\pi^2}G_\gamma(\lambda_{\chi_L})
\simeq \frac{G_F m_\mu^2}{8\sqrt{2}\pi^2}\left(\frac{5}{3}+O(\lambda_{\chi_L})\right),
\end{equation}
which entirely coincides with the result given in the PDG book \cite{ParticleDataGroup:2020ssz}.\\

Other contributions to the magnetic dipole moment by new physics under the considered scenario and involving light neutrino and W-boson interactions are very small and therefore ignorable. This fact is easy to figure out by looking at these three contributions, which are the last term in (\ref{FunAR}) and the two last terms in (\ref{FunAL}). The interference terms, which contain ratio $m_k/m_\mu$,  are strongly suppressed by a factor of $m_k/m_\mu\sim 10^{-6}$, where $m_k< 0.1{\rm eV}$ for light active neutrino mass and $m_\mu=106{\rm MeV}$ have been used; while the rest gets a tiny value due to being proportional to $U^{R\dagger}_{W_\mu}U^R_{W_\mu}\sim\tilde{R}_\nu^*\tilde{R}_\nu^T\ll U^{L\dagger}_{W_\mu}U^L_{W_\mu}$.\\
\begin{figure}[t]
\begin{center}
\includegraphics[width=7.5cm,height=5cm]{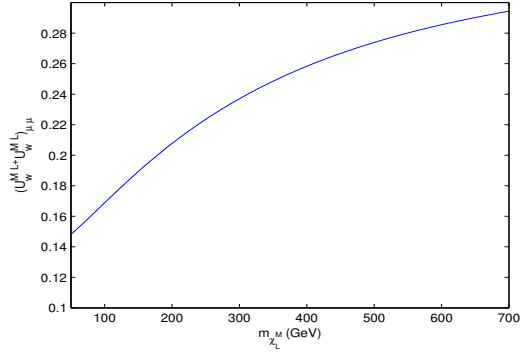}
\caption{The correlation between $\left(U_{W}^{ML\dagger}U_{W}^{ML}\right)_{\mu\mu}$ and heavy neutrino mass $m_\chi^M$ when muon anomalous magnetic dipole moment $\Delta a_\mu$ is set at its current best fit value; the channel of virtual W-boson and heavy 
neutrinos. 
}
\label{GaugePlot}
\end{center}
\end{figure}

In figure \ref{GaugePlot}, we show the correlation between $\left(U_{W}^{ML\dagger}U_{W}^{ML}\right)_{\mu\mu}$ and heavy neutrino mass $m_\chi^M$ when muon anomalous magnetic dipole moment is set at its current experimental best fit value 
$\Delta a_\mu =251\times 10^{-11}$, for the channels with participation of virtual W-boson and heavy 
neutrinos. The figure shows that the channel contribution is significant only if $\left(U_{W}^{ML\dagger}U_{W}^{ML}\right)_{\mu\mu}$ has magnitude about 0.1 or larger; however the real value is extremely tiny due to $U_{W}^{ML\dagger}U_{W}^{ML}\sim \tilde{R}_\nu^\dagger\tilde{R}_\nu\sim 10^{-12}$.\\

\begin{figure}[t]
\begin{center}
\includegraphics[width=7.5cm,height=5cm]{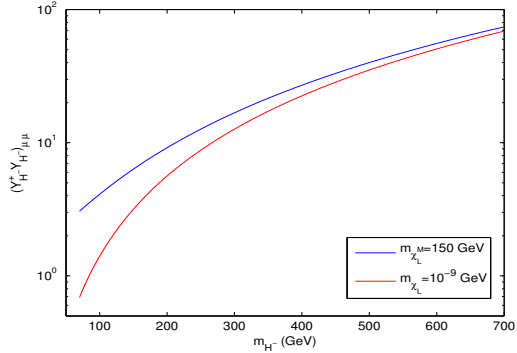}
\caption{ $(Y^\dagger_{H^-}Y_{H^-})_{\mu\mu}$ as function of physical singly charged Higgs scalar mass at some specific values of mirror neutrino masses; if only left or right sector is taken into account, $\Delta a_\mu$ is set at the present best fit value.}
\label{SingleH1Plot}
\end{center}
\end{figure}

\begin{figure}
\begin{center}
\begin{tabular}{cc}
\includegraphics[width=7cm,height=4.5cm]{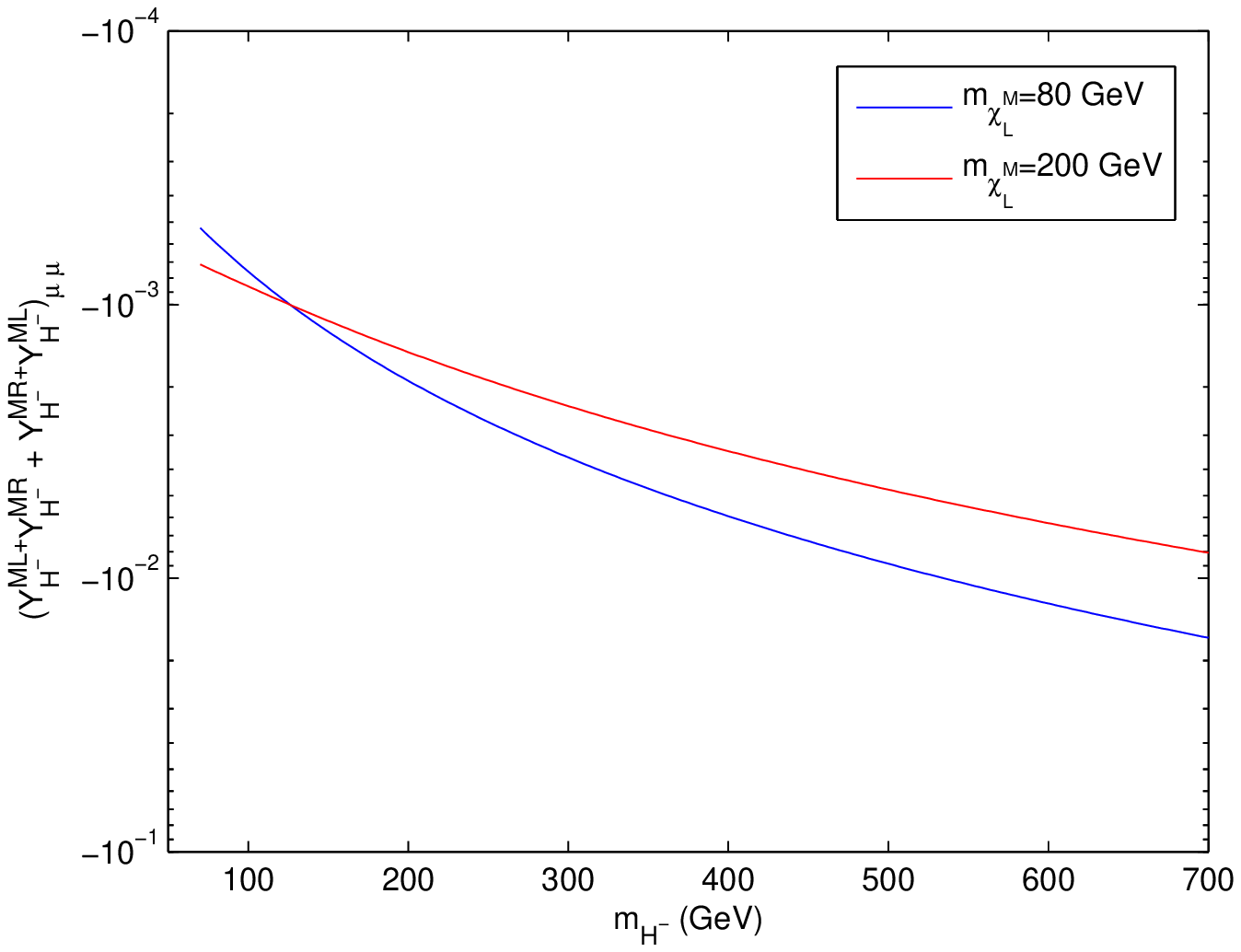}&
\includegraphics[width=7cm,height=4.5cm]{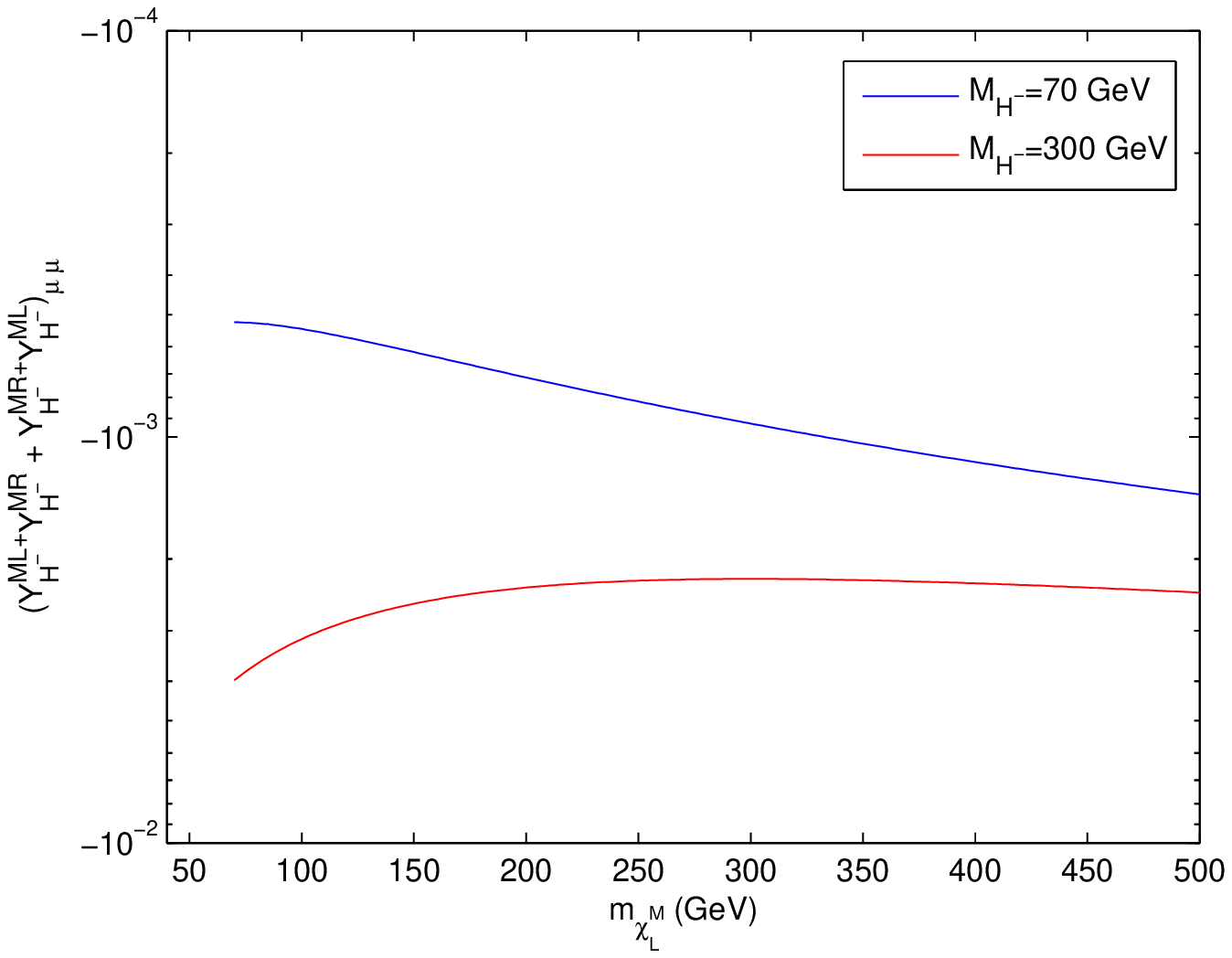}
\end{tabular}
\caption{The correlations between Yukawa couplings and : i, Singly charged Higgs mass (left-panel) for $m_{\chi^M_L}=80~(200)~{\rm GeV}$, blue (red) lines; ii, Heavy neutrino masses (right-panel) for $m_{H^-}=70~(300)~{\rm GeV}$, blue (red) lines, when $\Delta a_\mu=251\times 10^{-11}$ is fixed.}
\label{Contri_H1YrYL}
\end{center}
\end{figure}

We continue next with the contributions of 1-loop diagrams with virtual singly negative charged Higgs $H^-$. In contrast to the previously considered cases, the interference terms are enhanced by factor $m_k/m_\mu\sim 1000$, where $m_k\sim 100$ GeV and 
$m_\mu=106$ MeV; therefore strongly dominates in comparison with the others. To have a more intuitive understanding, we present the correlations between relevant Yukawa couplings as a function of neutrino masses of the channels in which virtual singly charged Higgs took part in the loops when muon anomalous dipole moment is fixed at the best-fit value. These correlations are presented for the cases if only the first term of (\ref{FunAR}) and (\ref{FunAL}) are taken into account (Fig.\ref{SingleH1Plot}) and all terms are considered (left-panel, Fig.\ref{Contri_H1YrYL}). The Yukawa coupling absolute values obtained in Fig.\ref{Contri_H1YrYL} are about three orders smaller than those in Fig.\ref{SingleH1Plot}, which are certainly consistent with the mentions in earlier parts. The negative sign of the horizontal axis in Fig.\ref{Contri_H1YrYL} shows that the interference terms, thus the channel, would give subtractions to the muon anomalous dipole moment if the involving Yukawa couplings are positive.\\ 
%
%

Ignoring the negative sign, the smallest absolute values of Yukawa couplings, which are easily figured out to be at $m_{H^-}=70$ GeV, obtained from Fig.\ref{Contri_H1YrYL} left-panel are
\begin{equation}
|((Y_{H^-}^{ML})^\dagger Y_{H^-}^{MR}+(Y_{H^-}^{MR})^\dagger Y_{H^-}^{ML})_{\mu \mu}|\simeq 5.29\times 10^{-4}(7.02\times 10^{-4}),
\end{equation}
for $m_{\chi^M_L}=80~(200)~{\rm GeV}$. These results are apparently available for both cases of $H_3^-$ and $H_{3M}^-$. 
Magnitude of $|((Y_{H^-}^{ML})^\dagger Y_{H^-}^{MR}+(Y_{H^-}^{MR})^\dagger Y_{H^-}^{ML})_{\mu \mu}|$, in fact, might be estimated basing on the model scheme and supposing that mirror charged lepton masses are about 100GeV, the calculation implies 
\begin{equation}
|((Y_{H^-}^{ML})^\dagger Y_{H^-}^{MR}+(Y_{H^-}^{MR})^\dagger Y_{H^-}^{ML})_{\mu \mu}|\sim 10^{-12}\times\left(\frac{s_M}{s_2c_M^2}\right)\frac{6m_\mu m_\ell^M}{M_W^2}\sim 7.0\times 10^{-15} \left(\frac{s_M}{s_2c_M^2}\right),
\end{equation}
for the case of $H^-_{3M}$; and $\sim 7.0\times 10^{-15} \left(\frac{s_M}{c_M}\right)^2$; if the negative scalar participating in loops is
$H^-_3$. Thus the real values are extremely tiny ($< 10^{-8}$), even if $s_2$ and $c_M$ of the mixing angles are as small as 0.01. 
The above analysis figure out a fact that contributions of  the singly charged Higgs scalar channel to the muon anomalous dipole moment 
are too small to be able to explain the muon anomalous magnetic dipole moment experimental results.\\   

\begin{figure}
\begin{center}
\begin{tabular}{cc}
\includegraphics[width=7cm,height=4.5cm]{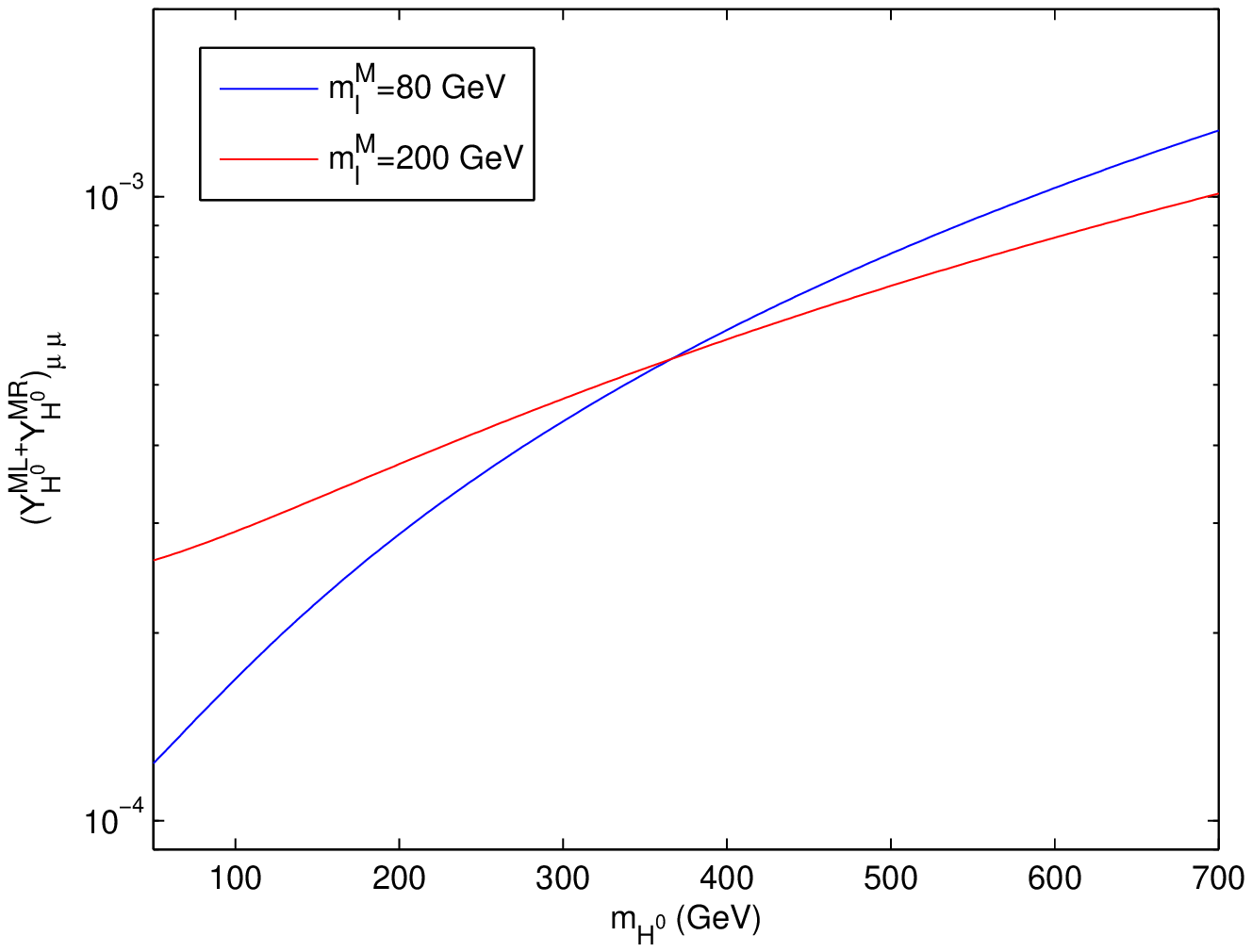}&
\includegraphics[width=7cm,height=4.5cm]{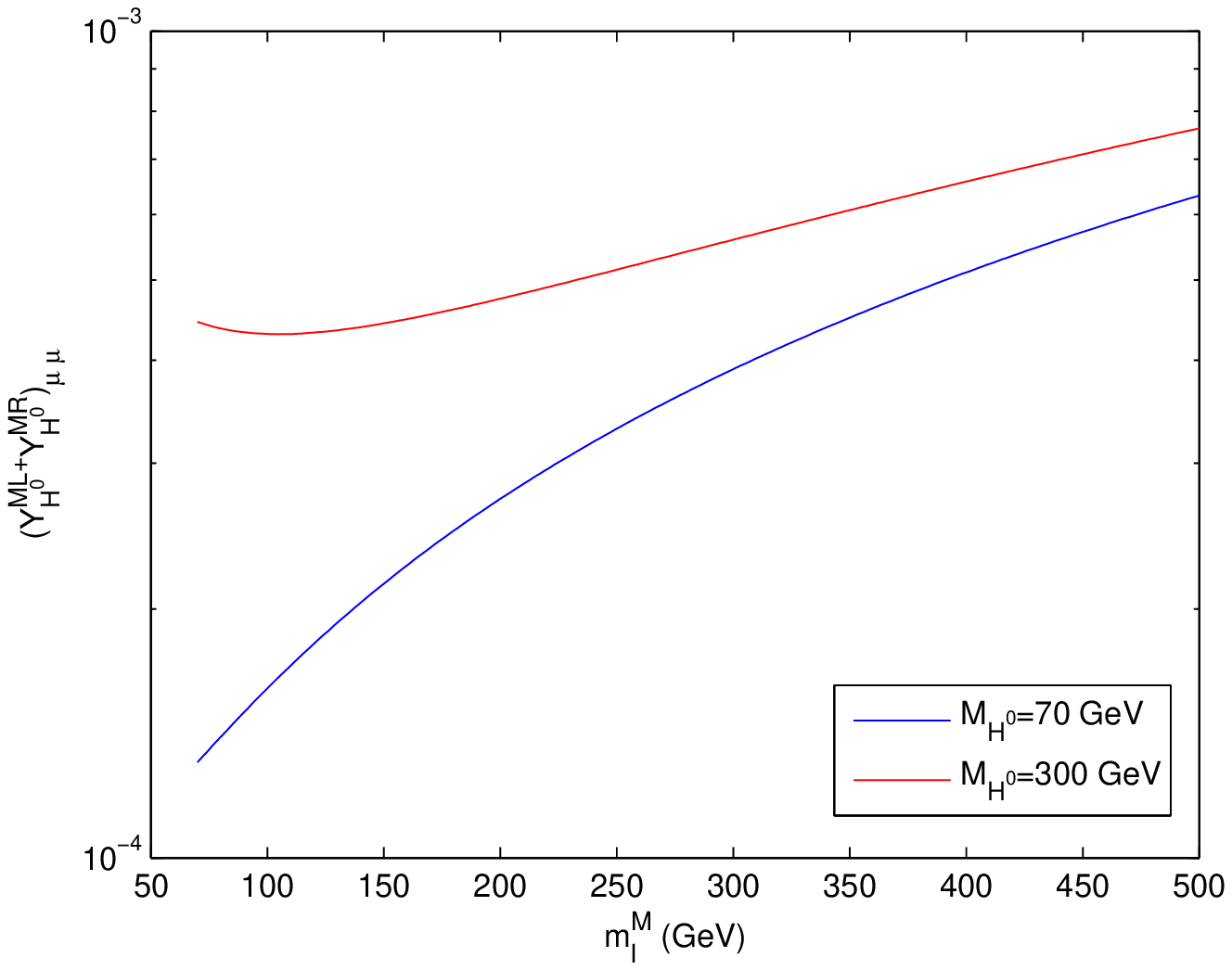}
\end{tabular}
\caption{The correlations between Yukawa couplings and : i, Heavy neutral Higgs scalar masses (left-panel); ii, Mirror charged lepton masses (right-panel) for diagrams, whose particles running inside loops are heavy physical scalars and mirror charged leptons, $\Delta a_\mu=251\times 10^{-11}$ is fixed.}
\label{Contri_H0YrYL}
\end{center}
\end{figure}

The correlations between the Yukawa couplings involving the heavy neutral Higgs scalars (which are ${\tilde{H}}_i^0$, 
(i=1,2,3), $H_3^0$ and $H_{3M}^0$) and either the Higgs or mirror charged lepton masses are respectively shown in the Fig.\ref{Contri_H0YrYL} 
 left (or right) panel. The discussion on light Higgs channel will be presented in a latter separate part due to the enormous difference in mass hierarchies. The same as previous considered case of $H^-$ channels, contributions of the diagrams involving neutral scalar 
(including both light and heavy ones) diagrams are predominated by the mixing terms due to the heaviness in masses of mirror 
charged leptons. To explain the muon anomalous dipole moment, magnitude of $(Y_{H^0}^{ML\dagger}Y_{H^0}^{MR})_{\mu\mu}$ requires a value
within $10^{-4}-10^{-3}$ range, which slightly increase with the increasing of neutral Higgs scalar mass (Fig.\ref{Contri_H0YrYL}, left panel). At the initial points of the lines, corresponding to $m_{H^0}\simeq 50$GeV, we have  
\begin{equation}
(Y_{H^0}^{ML\dagger}Y_{H^0}^{MR})_{\mu\mu}\simeq 1.28\times 10^{-4}(2.61\times 10^{-4}),
\end{equation}
for $m_{m_\ell^M}=80~(200)~{\rm GeV}$, respectively. Carry on the same strategy as in earlier part, the theoretical estimation can be performed using eqs. from (\ref{YuV01}) to (\ref{YuV06}). The results arrive at  
\begin{equation}
\label{YYH0}
(Y_{H^0}^{ML\dagger}Y_{H^0}^{MR})_{\mu\mu}\sim 2\alpha^2\frac{\left(\tilde{R}_\ell^\dagger(m_{\ell M}^d)^2\tilde{R}_\ell\right)_{\mu\mu}}{M_W^2}\sim 4.69\alpha^2\times 10^{-12},
\end{equation}
where mirror charged lepton masses are taken about $100$GeV, $\alpha$ denotes for $\frac{\alpha_i}{s_2}$, $\frac{s_M}{c_M}$ or $\frac{s_2}{s_{2M}}$, corresponding to ${\tilde{H}}_i^0$, 
(i=1,2,3), $H_3^0$ or $H_{3M}^0$, respectively. Notes that to obtain eq. (\ref{YYH0}), first terms of the Yukawa couplings defined from 
(\ref{YuV01}) to (\ref{YuV06}), which contain $m_\ell^d$ thus sub-dominate in comparison to the second ones with $m_{\ell M}^d$, are reasonably neglected. Equation (\ref{YYH0}) also means the currently considered channels might provide contributions to the muon anomalous magnetic dipole moment about three orders higher than those of the singly charged scalars in magnitude. In fact, real values of charged lepton masses can be larger at order of several hundreds GeV. For instance, if $m_\ell^M=500$ GeV is taken, $(Y_{H^0}^{ML\dagger}Y_{H^0}^{MR})_{\mu\mu}\sim 1.17\alpha^2\times 10^{-10}\sim 1.12\times 10^{-6}$ for $\alpha=100$, which occurs at $s_2=0.01$, $c_M=0.01$, or $s_{2M}=0.01$, corresponding to the case of ${\tilde{H}}_i^0$, (i=1,2,3), $H_3^0$ or $H_{3M}^0$, respectively. Therefore, contributions of the heavy neutral Higgs scalar 
channels can not be able to explain the muon anomalous magnetic dipole moment, but they might be possible to provide sizable corrections.\\
\begin{figure}
\begin{center}
\begin{tabular}{cc}
\includegraphics[width=7cm,height=4.5cm]{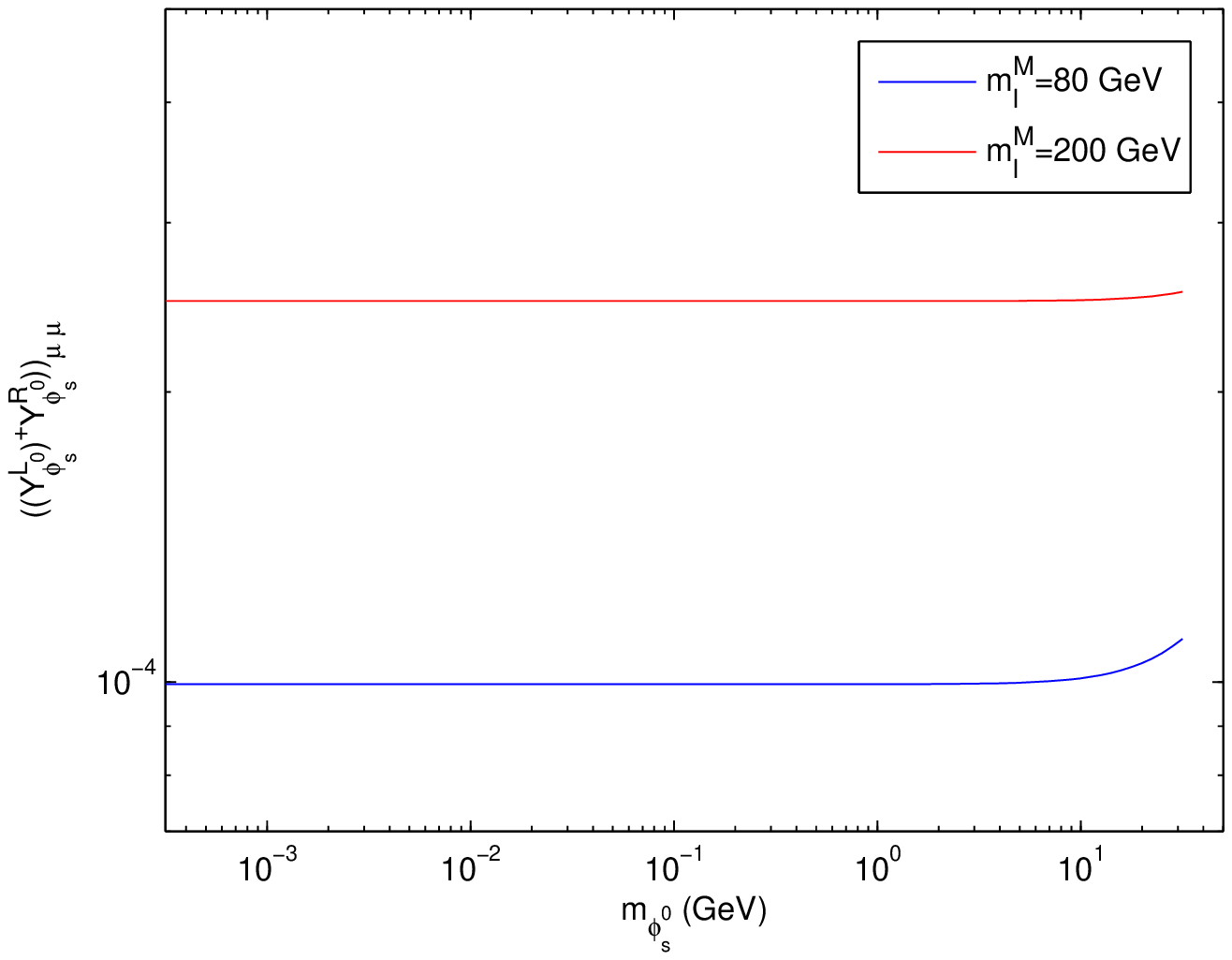}&
\includegraphics[width=7cm,height=4.5cm]{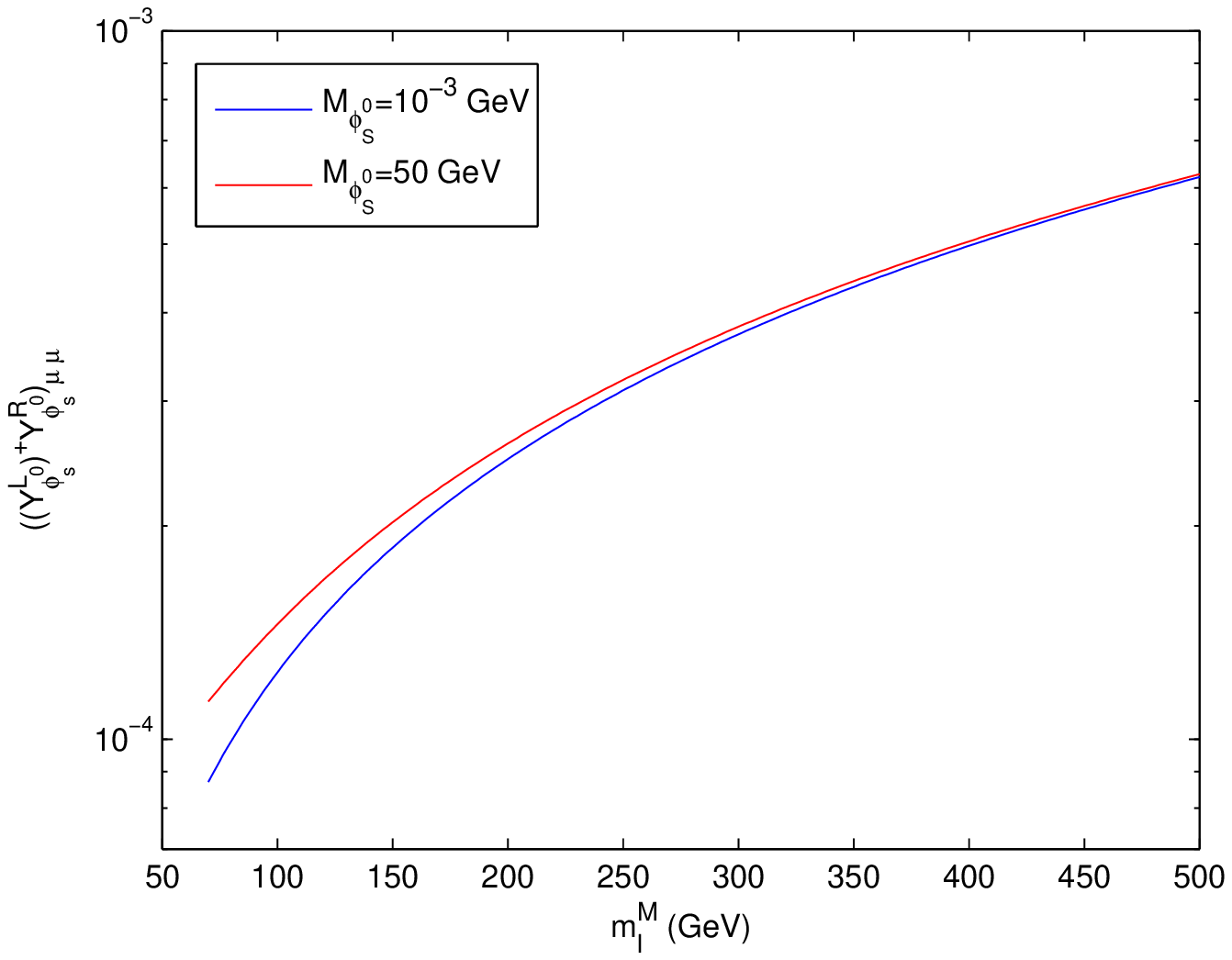}
\end{tabular}
\caption{The same content as Fig.\ref{Contri_H0YrYL}, for the case of diagrams with loops formed by light Higgs scalar and 
mirror charged leptons.}
\label{Contri_Phi0YrYL}
\end{center}
\end{figure}

The same contents as Fig.\ref{Contri_H0YrYL} are presented in Fig.\ref{Contri_Phi0YrYL}, in which heavy neutral Higgs masses are replaced by those of the light one at lower scale from keV to order of GeV. At a given value of $m_\ell^M$, magnitude of $((Y^L_{\phi_s^0})^\dagger Y^R_{\phi_s^0})_{\mu\mu}$ does not change with the increase in light Higgs scalar mass until about 10 GeV, then slowly increase. From Fig.\ref{Contri_H0YrYL} left panel, we easily obtain from the constant lines 
\begin{equation}
\label{YYPhi0}
((Y^L_{\phi_s^0})^\dagger Y^R_{\phi_s^0})_{\mu\mu}\simeq 8.75\times 10^{-5}(1.13\times 10^{-4}),
\end{equation}
corresponding to $m_{m_\ell^M}=80~(200)~{\rm GeV}$, respectively. After a simple computation, the result arrives at relations of Yukawa coupling matrix
\begin{equation}
\label{ggPhi0}
|\tilde{g}_{\ell s}^\dagger \tilde{g}_{\ell s}|_{\mu\mu}\simeq 5.39\times 10^{-5}(6.96\times 10^{-5}),
\end{equation}
which are five orders larger than the upper constraints obtained by current experimental bound of $\mu\to e \gamma$ decay \cite{Dinh:2021fmt}\footnote{Here we have recast eq. (51) in \cite{Dinh:2021fmt} to have similar form as (\ref{ggPhi0}) for more convenient in comparison.}
\begin{equation}
\label{ggPhi0Mu2E}
|\tilde{g}_{\ell s}^\dagger \tilde{g}_{\ell s}|_{\mu e}\simeq 5.29\times 10^{-10}(1.30\times 10^{-9}).
\end{equation}
The two equations (\ref{ggPhi0}) and (\ref{ggPhi0Mu2E}) are easily concurrently fulfilled; if $\tilde{g}_{\ell s}$ is proportional to a matrix, which is close to unitary form. Beside being constrained by muon anomalous dipole moment and the rare lepton flavor violation 
decay $\mu\to e\gamma$, $g_{\ell s}$ is also constrained by searching new particles at the LHC, especially new scalars in the high mass
region. The previous publication \cite{Hoang:2014pda} studied the decay $H_3^0\rightarrow \bar{\ell}^M\ell^M\rightarrow \bar{\ell}\phi_s\ell\phi_s^*$, where $\phi_s$ is invisible and considered as missing transverse energy. This process
is identified with 2 leptons and missing energies, which imitates the signal of scalar decay $H_{SM}^0\rightarrow W^-W^+\rightarrow \bar{\ell}\nu\ell\bar{\nu}$. No excess over the background was detected at both ATLAS and CMS, however an upper constraint $|g_{\ell s}|^2\le 10^{-6}$
was obtained (see \cite{Hoang:2014pda}). This limit is smaller than the required values obtained in (\ref{ggPhi0}), therefore it disfavors the possibility to explain the discrepancy of the muon magnetic dipole moment with the standard model prediction. For more quantitative view, if $|g_{\ell s}|^2= 10^{-6}$ is taken, the channel of light 
neutral scalar might contribute to the muon anomalous magnetic dipole moment a deviation as
\begin{equation}
\Delta a_\mu({\phi_s^0})\simeq 4.7\times 10^{-11},
\end{equation}   
which is about size of the uncertainty of the two-loop electroweak contribution and smaller than those of the hadronic effects. 
Therefore, the contribution of this model to the muon problem is currently not important. However, it might have a role in the future when 
we have more precise measurements on the muon magnetic dipole moment. 

\section{\label{conl} Conclusion}
Recent experimental measurements on muon magnetic dipole moment show substantial discrepancies with 
standard model predictions. These differences might be understood in scenarios of physics beyond the standard model with new particles and interactions. In this work, we have derived an algebraic formula and performed numerical analysis for 
muon anomalous magnetic dipole moment at 1-loop approximation in an extended model with mirror symmetry, in which the light neutrino 
masses are generated by the type-I see-saw at low scale of the electroweak. We have shown that the contributions provided by neutrino and either W boson or singly charged Higgs scalar channels are too small to be taken into account. Moreover, the channel of heavy neutrinos and singly charged Higgs also gives a contribution opposite to that of the rest. For the case that particles running inside loops are heavy neutral Higgs scalars and mirror charged leptons, although contribution of this channel is not able to explain the experimental results, it might provide a sizable amount of correction to the muon anomalous magnetic dipole moment; if at least one of the following quantities $s_2$, $c_M$, or $s_{2M}$ has the value as small as 0.01 or less. As the most promising case, the channel involving light neutral Higgs 
scalar is expected to be able to explain the muon anomalous problem, however the required magnitude of $g_{\ell s}$ is larger than its upper
limit obtained from experimental search on heavy new scalars. In any case, the contribution of this channel might become more important in the 
future when the uncertainties of the two-loop electroweak and hadronic contributions are improved.

\section*{Acknowledgments}
	
This research is funded by Vietnam National Foundation for Science and Technology Development (NAFOSTED) 
under grant number 103.01-2019.307.
	
\bibliography{combine}
\end{document}